\documentclass[12pt,superscriptaddress,aps,prd,preprint]{revtex4}
\usepackage{amsmath}
\usepackage{amssymb}
\usepackage{graphicx}
\usepackage{url}
\makeatletter

\newcommand{\bea}{\begin{eqnarray}}
\newcommand{\eea}{\end{eqnarray}}

\begin{document}
\title{ On Friedmann-Robertson-Walker Model in Conformal Teleparallel Gravity}

\author{J. G. Silva}\email[]{juceliags@gmail.com}
\affiliation{Instituto de F\'{i}sica and International Center of Condensed Matter Physics,
Universidade de Bras\'{i}lia, 70910-900, Bras\'{i}lia, DF,
Brazil.}

\author{A. F. Santos}\email[]{alesandroferreira@fisica.ufmt.br}
\affiliation{Instituto de F\'{\i}sica, Universidade Federal de Mato Grosso,\\
78060-900, Cuiab\'{a}, Mato Grosso, Brazil.\\
Department of Physics and Astronomy, University of Victoria,
3800 Finnerty Road Victoria, BC, Canada.}

\author{S. C. Ulhoa}\email[]{sc.ulhoa@gmail.com}
\affiliation{Instituto de F\'{i}sica and International Center of Condensed Matter Physics,
Universidade de Bras\'{i}lia, 70910-900, Bras\'{i}lia, DF,
Brazil.}

\begin{abstract}
In this paper we use the conformal teleparallel gravity to study an isotropic and homogeneous Universe which is settled by the Friedmann-Robertson-Walker (FRW) metric. The conformal symmetry demands the existence of a scalar field which works as a dark field for this model. We solve numerically the field equations then we obtain the behavior of some cosmological parameters such as the scale factor, the deceleration parameter and the energy density of the perfect fluid which is the matter field of our model. The field equations, that we called modified Friedmann equations, allow us to define a dark fluid, with dark energy density and pressure, responsible for the acceleration in the Universe, once we defined an equation of state for the dark fluid.
\end{abstract}

\maketitle

\section{Introduction}

The accelerated expansion of the universe is a cosmological phenomenon that has recently been confirmed by several observational data, such as, Supernovae Ia \cite{Perlm}, large scale structure \cite{Tegmark}, the baryon acoustic oscillations \cite{Eisenstein}, cosmic microwave background radiation \cite{Spergel}, and weak lensing \cite{Jain}. In order to explain this phenomenon, there are two ways which are being investigated in the literature. The first introduces an exotic fluid, called dark energy, with negative pressure in the theory of general relativity. The second possibility consists in modifying  general relativity which is known as modified theories of gravity. For instance we can cite $f(R)$ theories and Brans-Dicke gravitation~\cite{MG}.

Among the various theories of modified gravity, here we will use teleparallel gravity as formulated in \cite{Maluf1, Maluf2}. The main advantage of using teleparallel gravity is that in its framework it is possible to define well-behaved expressions for energy, momentum and angular momentum of the gravitational field \cite{maluf:335,Ulhoa, maluf21}. Therefore it seems natural to generalize such a theory to include the conformal invariance which is a fundamental symmetry of spacetime. Scale invariance is a desirable feature of physical properties, thus conformal invariance can be understood as a generalization of it. It should be noted that in this case the conformal factor should depend on the coordinates. In order to understand what is a conformal transformation we consider a spacetime ($\cal{M}$, $g_{ab}$), where $\cal{M}$ is a smooth n-dimensional manifold and $g_{ab}$ is a metric in $\cal{M}$. The following conformal transformation
\bea
\tilde{g}_{ab}=e^{2\theta(x)}g_{ab}(x), \label{eq1}
\eea
where $e^{2\theta(x)}$ is a smooth non-vanishing function of the coordinates, is a point-dependent rescaling of the metric. Thus $e^{2\theta(x)}$ is called a conformal factor. It is well-known that such transformations preserve angles and change distances between two points which are described by the same coordinate system. Scale transformations are recovered when $\theta=const.$ A field theory which is invariant under these transformations is named a conformal field theory (CFT), i.e., the physics of the theory looks the same at all length scales.

The conformal transformation given by equation (\ref{eq1}) can have different interpretations depending on whether the metric is a fixed background metric, or a dynamical background metric. If the metric is dynamical background, the transformation is a diffeomorphism; this is a gauge symmetry, and if the background is fixed, the transformation should be thought of as a physical symmetry, taking the point $x^\alpha$ to point $\tilde{x}^\alpha$. A review of conformal field theory can be seen in \cite{Matthias} and several details in the textbooks \cite{Haw, Fujii, Carroll}. In recent years, several works involving conformal transformations have been developed. For example, in \cite{Cham1, Cham2, Haw2, Arm} an inflationary cosmology was studied in this context; the phenomenon of asymptotic conformal invariance was analyzed in \cite{Buch1, Buch2}; gravitational theories including higher order terms of the curvature tensor with respect to the conformal symmetry have been studied in \cite{Man1, Man2, Moon}, while in \cite{Hooft1, Hooft2} it is discussed the possibility of a conformal theory of gravity being a candidate for quantum gravity.

There are at least two ways to write a gravitational theory invariant under conformal transformations. The first way is choosing a lagrangian with the quadratic term of Weyl tensor and the second one is adding a scalar field $\phi$ to the Hilbert-Einstein action, together with a suitable kinetic term for the scalar field. In this paper we will investigate the second option, following the theory presented in \cite{Maluf1} where teleparallel gravity with conformal symmetry was constructed. Recently, this gravitational model with conformal invariance is the subject of several investigations. For instance, in \cite{Formiga} it is shown an equivalence between the conformally invariant teleparallel theory and a particular Weyl Teleparallel theory, in \cite{Bamba} it is demonstrated that, in extended teleparallel gravity with a conformal scalar field, a power-law of the de Sitter expansions of the universe can occur. In \cite{Myr}  some cosmological aspects of conformal teleparallel gravity are also studied. Thus Conformal Teleparallel gravity has been extensively used to understand cosmological models.

The structure of this paper is as follows. In the next section we analyze the construction of conformal teleparallel gravity. In the Section \ref{frw} we will investigate the FRW metric in this gravitational model and we solve the field equations. We conclude our discussion in section \ref{con}.

{\bf Notation:} Spacetime indices $\mu, \nu, \rho,...$ run from 0 to 3, where $(\mu=0),(i=1,2,3)$ are the time coordinate and space of spacetime, $a, b, c, ...$ run from 1 to 3 or $(0),(i)$ are the indices representing $SO(3,1)$. In addition, we adopt units where $G=\hbar=c=1$, unless otherwise stated.

\section{Conformal Teleparallel Gravity}\label{ct}

Teleparallelism Equivalent to General Relativity (TEGR) is a well-defined theory of gravitation. It started with Einstein himself in a attempt to unify electromagnetism and gravitation~\cite{einstein}. TEGR is formulated in the framework of
Weitzenb\"ock geometry which is endowed with the Cartan connection~\cite{Cartan}, $\Gamma_{\mu\lambda\nu}=e^{a}\,_{\mu}\partial_{\lambda}e_{a\nu}$,
where $e^{a}\,_{\mu}$ is the tetrad field. It is similar to what happens in the riemannian geometry which is endowed with the Christofell symbols ${}^0\Gamma_{\mu \lambda\nu}$. However the dynamical variables are the tetrads rather than the metric tensor components.  The tetrad field is related to the metric tensor by means the well known relation $g_{\mu\nu}=e^a\,_\mu e_{a\nu}$. In the riemannian geometry the manifold is characterized by the curvature and has a vanishing torsion, on the other hand in the Weitzenb\"ock geometry the manifold is characterized by the torsion, $T^{a}\,_{\lambda\nu}=\partial_{\lambda} e^{a}\,_{\nu}-\partial_{\nu}
e^{a}\,_{\lambda}$, and has a vanishing curvature. Thus both descriptions seems to be opposite, but they are in fact equivalent. Such equivalence is settled by the following mathematical identity
\begin{equation}
\Gamma_{\mu \lambda\nu}= {}^0\Gamma_{\mu \lambda\nu}+ K_{\mu
\lambda\nu}\,, \label{2}
\end{equation}
where
\begin{eqnarray}
K_{\mu\lambda\nu}&=&\frac{1}{2}(T_{\lambda\mu\nu}+T_{\nu\lambda\mu}+T_{\mu\lambda\nu})\label{3.5}
\end{eqnarray}
is the contortion tensor.
Thus one could calculate the scalar curvature from Christofell symbols in terms of the torsion of the Weitzenb\"ock geometry using the expression (\ref{2}). It reads
\begin{equation}
eR(e)\equiv -e({1\over 4}T^{abc}T_{abc}+{1\over
2}T^{abc}T_{bac}-T^aT_a)+2\partial_\mu(eT^\mu)\,,\label{5}
\end{equation}
where $e$ is the determinant of the tetrad field, $T_a=T^b\,_{ba}$ and
$T_{abc}=e_b\,^\mu e_c\,^\nu T_{a\mu\nu}$. Therefore there is an analogue of the Hilbert-Einstein lagrangian density, which is defined in the riemannian geometry, in the Weitzenb\"ock space-time. This is the TEGR lagrangian density, it is given by
\begin{equation}
\mathfrak{L}(e_{a\mu})=-ke\left(\frac{1}{4}T^{abc}T_{abc}+\frac{1}{2}T^{abc}T_{bac}-T^{a}T_a\right)\,,\label{lagran}
\end{equation}
once the total divergence is dropped out, since it doest not alter the field equations. The coupling constant is  $k=\frac{1}{16\pi}$. Hence a theory obtained from such a lagrangian density is dynamically equivalent to General Relativity. However it is not the same theory since it is possible to define a gravitational energy-momentum tensor in the realm of TEGR \cite{maluf:335,PhysRevLett.84.4533} which cannot be done in General Relativity.

The choice of the tetrad field establishes the observer which is true even in the context of the tetrad formulation of General Relativity \cite{Landau,Maluf:2007qq}. Such a feature settles down the issue about the degrees of freedom of the tetrad field. The metric tensor has 10 independent components, due to its symmetry, while the tetrad field has 16 components. These 6 remaining components are established by the reference frame choice. Thus for each metric we have infinite possible tetrad fields.

It should be noted that the lagrangian density (\ref{lagran}) can be rewritten as
\begin{equation}
\mathfrak{L}(e_{a\mu})=-ke\Sigma^{abc}T_{abc}\,,
\end{equation}
where
\begin{equation}
\Sigma^{abc}=\frac{1}{4}\left(T^{abc}+T^{bac}-T^{cab}\right)+ \frac{1}{2}\left(\eta^{ac}T^{b}-\eta^{ab}T^{c}\right)\,.
\end{equation}
We notice that it is not invariant under conformal transformations since $g_{\mu\nu} \rightarrow \widetilde{g}_{\mu\nu}= e^{2\theta (x)}g_{\mu\nu}$, $ e_{a\mu} \rightarrow \widetilde{e}_{a\mu}=e^{\theta (x)}e_{a\mu}$ and $e \rightarrow \widetilde{e}=e^{4\theta}e$. Here the parameter $\theta=\theta(x)$ is arbitrary.

Therefore in order to have a conformal teleparallel gravity we need an invariant lagrangian density. Maluf showed that the above lagrangian density should be modified to accomplish such a task. Here we follow the proposition presented by Maluf in \cite{Maluf1}. Thus the complete lagrangian density invariant under conformal transformations is given by
\begin{eqnarray}
\mathfrak{L}(e_{a\mu},\phi)=ke\big[-\phi^{2}\Sigma^{abc}T_{abc}+ 6 g^{\mu\nu}\partial_{\mu}\phi \partial_{\nu}\phi - 4g^{\mu\nu}\phi(\partial_{\mu}\phi)T_{\nu}\big]+\mathfrak{L}_m\,,\label{cl}
\end{eqnarray}
where $\phi$ is a scalar field and $\mathfrak{L}_m$ is the lagrangian density of the matter fields. We point out that the transformation property of the scalar field, given by $\phi \rightarrow \widetilde{\phi}=e^{-\theta}\phi$, leaves invariant the above lagrangian density.

If we perform a variation of the lagrangian density (\ref{cl}) with respect to $\phi$, then the field equation reads
\begin{eqnarray}
\partial_{\nu}(eg^{\mu\nu}\partial_{\mu}\phi)+\frac{1}{6}\phi\left[e\Sigma^{abc}T_{abc}-2\partial_{\mu}(eT^{\mu})\right]=\frac{1}{12k}\frac{\delta{\mathfrak{L}_{m}}}{\delta {\phi}}\,,
\end{eqnarray}
which can be rewritten as
\begin{eqnarray}
\partial_{\nu}(eg^{\mu\nu}\partial_{\mu}\phi)-\frac{1}{6} e\phi R(e)=\frac{1}{12k}\frac{\delta{\mathfrak{L}_{m}}}{\delta {\phi}}\,,\label{eqp}
\end{eqnarray}
once we use the relation $\left[e\Sigma^{abc}T_{abc}-2\partial_{\mu}(eT^{\nu})\right]\equiv -e R(e)$.

Similarly if we perform a variational derivative of the lagrangian density (\ref{cl}) with respect to the tetrad field $e_{a\mu}$, it will yield the following field equations
\begin{eqnarray}
&&\partial_{\lambda}(e\phi^2\Sigma^{a\mu\lambda})-e\phi^2\big(\Sigma^{b\lambda\mu}T_{b\lambda}^{\hspace*{0.2cm} a}-\frac{1}{4}e^{a\mu}\Sigma^{bcd}T_{bcd}\big) -  \frac{3}{2}
e e^{a\mu}g^{\sigma\nu}\partial_{\sigma}\phi \partial_{\nu}\phi +\nonumber\\&& 3e e^{a\nu}g^{\sigma\mu}\partial_{\sigma}\phi \partial_{\nu}\phi + ee^{a\mu}g^{\sigma\nu}T_{\nu}\phi(\partial_{\sigma}\phi)-e\phi e^{a\sigma}g^{\mu\nu}\left(T_{\nu}\partial_{\sigma} \phi+T_{\sigma}\partial_{\nu}\phi\right) -\nonumber\\&& eg^{\sigma\nu}\phi(\partial_{\sigma}\phi)T^{\mu a}_{\hspace{0.2cm} \nu} - \partial_{\rho}\left[eg^{\sigma\mu}\phi(\partial_{\sigma}\phi) e^{a\rho} \right] + \partial_{\nu}\left[eg^{\sigma\nu}\phi(\partial_{\sigma}\phi) e^{a\mu} \right]=\frac{1}{4k}\frac{\delta{\mathfrak{L}_{m}}}{\delta {e_{a\mu}}}\,,\label{eqt}
\end{eqnarray}
where
\begin{equation*}
\frac{\delta {\mathfrak{L}_{m}}}{\delta {e_{a\mu}}}=ee^{a}\,_{\nu}T^{\nu\mu}\,.
\end{equation*}
We note that these equations reduce to those of TEGR when $\phi=1$. In addition if we take the trace of equation (\ref{eqt}) and combine with equation (\ref{eqp}) then it yields
\begin{equation}
\phi\frac{\delta{\mathfrak{L}_{m}}}{\delta {\phi}}=eT\,,
\end{equation}
where $T=g_{\mu\nu}T^{\nu\mu}$ is the trace of the energy-momentum tensor. Hence a traceless energy-momentum tensor just vanishes the right-side of equation (\ref{eqp}). Such a feature is necessary in a conformal theory to preserve the symmetry, as it happens in electromagnetism, and the continuity equation for instance when applied to a perfect fluid in the context of conformal Einstein equations~\cite{CEM}.


\section{Isotropic and Homogenous Universe}\label{frw}
In this section we'll analyze an isotropic and homogeneous Universe in the context of conformal teleparallel gravity. It is settled by the Friedmann-Robertson-Walker (FRW) metric which is given by
\begin{equation}
ds^2=-dt^2+a^2(t)\left[\frac{dr^2}{(1-kr^2)}+r^2d\theta^2+r^2\sin^2\theta d\phi'^2\right],
\end{equation}
where $a(t)$ is the scale factor and $k$ assumes the values $(-1,0,1)$ depending on the features of the Universe we want to work with.
Let us choose a reference frame adapted to a stationary observer, thus the tetrad field is given by
\begin{equation}
e_{a\mu}=\left(
  \begin{array}{cccc}
    -1 & 0 & 0 & 0 \\
    0 & \frac{a}{(1-kr^2)^{\frac{1}{2}}} & 0 & 0 \\
    0 & 0 &  ar & 0 \\
    0 & 0 & 0 & a r\sin\theta \\
  \end{array}
\right)\,.
\end{equation}

We calculate the components of the tensors $T^{abc}=e^{b\mu}e^{c \nu}T^a\,_{\mu\nu}$ and $\Sigma^{abc}$ in order to obtain the field equations. It's worth recalling that these components are skew-symmetric in the last two indices, thus the non-vanishing components $T^{abc}$ read
\begin{eqnarray}
T^{(1)(1)(0)}&=&T^{(2)(2)(0)}=T^{(3)(3)(0)}=\frac{\dot{a}}{a}\,,\nonumber\\
T^{(2)(2)(1)}&=&T^{(3)(3)(1)}=-\frac{\small{(1-kr^2)^{\frac{1}{2}}}}{ar}\,,\nonumber\\
T^{(3)(3)(2)}&=&-\frac{\cot \theta}{ar}\,,
\end{eqnarray}
and the non-vanishing components $\Sigma^{abc}$ are given by
\begin{eqnarray}
\Sigma^{(0)(0)(1)}&=&-\frac{(1-kr^2)^{\frac{1}{2}}}{ar}\,,\nonumber\\
\Sigma^{(1)(1)(0)}&=&\Sigma^{(2)(2)(0)}=\Sigma^{(3)(3)(0)}=-\frac{\dot{a}}{a}\,,\nonumber\\
\Sigma^{(0)(0)(2)}&=&\Sigma^{(1)(2)(1)}=-\frac{1}{2}\frac{\cot\theta}{ar}\,,\nonumber\\
\Sigma^{(2)(2)(1)}&=&\Sigma^{(3)(3)(1)}=\frac{1}{2}\frac{(1-kr^2)^{\frac{1}{2}}}{ar}\,.
\end{eqnarray}
In addition we need the components $T_\mu$ which can be written in the following form
$T_\mu=(-\frac{3\dot{a}}{a};-\frac{2}{r};-\cot\theta;0)$. Therefore we get

$$\Sigma^{abc}T_{abc}= 6\left(\frac{\dot{a}}{a}\right)^2-2\frac{(1-kr^2)}{a^2r^2}\,.$$

We must work with a traceless energy-momentum tensor. Thus if we assume that $\phi=\phi(t)$ then the equation (\ref{eqp}) reads
\begin{eqnarray}
\ddot{\phi} + 3 \left(\frac{\dot{a}}{a}\right)\dot{\phi}+ \left[\frac{\ddot{a}}{a}+\left(\frac{\dot{a}}{a}\right)^2+\frac{k}{a^2}\right]\phi=0 \,.
\end{eqnarray}
In equation (\ref{eqt}) we choose the conformal perfect fluid energy-momentum tensor which is given by
\begin{equation}
T^{\mu\nu}=(\tilde{\rho}+\tilde{p})U^{\mu}U^{\nu}+ \tilde{p}g^{\mu\nu}\,.
\end{equation}
Thus the feature of trace-free yields the equation of state $\tilde{\rho}=3\tilde{p}$, since $U^{\mu}U_{\mu}=-1$.

We have just two independent components of eq. (\ref{eqt}), they are for $\mu=0$ and $a=(0)$
\begin{equation}
3\left[\frac{k}{a^2}+\left(\frac{\dot{a}}{a}\right)^2\right]+3\left(\frac{\dot{\phi}}{\phi^2}\right)\left[\dot{\phi}+2\phi\left(\frac{\dot{a}}{a}\right)\right]=8\pi \rho\,,\label{00}
\end{equation}
where $\rho=\tilde{\rho}/\phi^2$,
and for $\mu=1$ and $a=(1)$
\begin{equation}
- \left[2\left(\frac{\ddot{a}}{a}\right)+\left(\frac{\dot{a}}{a}\right)^2+\frac{k}{a^2}\right]-4\left(\frac{\dot{\phi}}{\phi}\right)\left(\frac{\dot{a}}{a}\right)+\left(\frac{\dot{\phi}}{\phi}\right)^2-2\left(\frac{\ddot{\phi}}{\phi}\right)=8\pi p\,,\label{11}
\end{equation}
where $p=\tilde{p}/\phi^2$. The above equations are the modified Friedmann equations. It should be noted that the conformal energy and pressure are related to the quantities of the ordinary perfect fluid.

In view of equation (\ref{00}) we can define an energy density whose source is the scalar field which is necessary to establish the conformal invariance in the theory. This is given by
\begin{equation}
\rho_{D}=\frac{1}{8\pi}\left[3\left(\frac{\dot{\phi}}{\phi}\right)^2+ 6\left(\frac{\dot{\phi}}{\phi}\right) \left(\frac{\dot{a}}{a}\right)\right]
\end{equation}
and from equation (\ref{11}) we have an extra pressure due to the scalar field, it reads
\begin{equation}
p_{D}=\frac{1}{8\pi}\left[-2\left(\frac{\ddot{\phi}}{\phi}\right)+ \left(\frac{\dot{\phi}}{\phi}\right)^2- 4\left(\frac{\dot{\phi}}{\phi}\right) \left(\frac{\dot{a}}{a}\right)\right].
\end{equation}
These extra terms are responsible by an acceleration of the Universe, in other words they have the same features of the so called dark energy. Thus we associate them to a dark fluid.

\begin{figure}[!htb]
\includegraphics[scale=0.5]{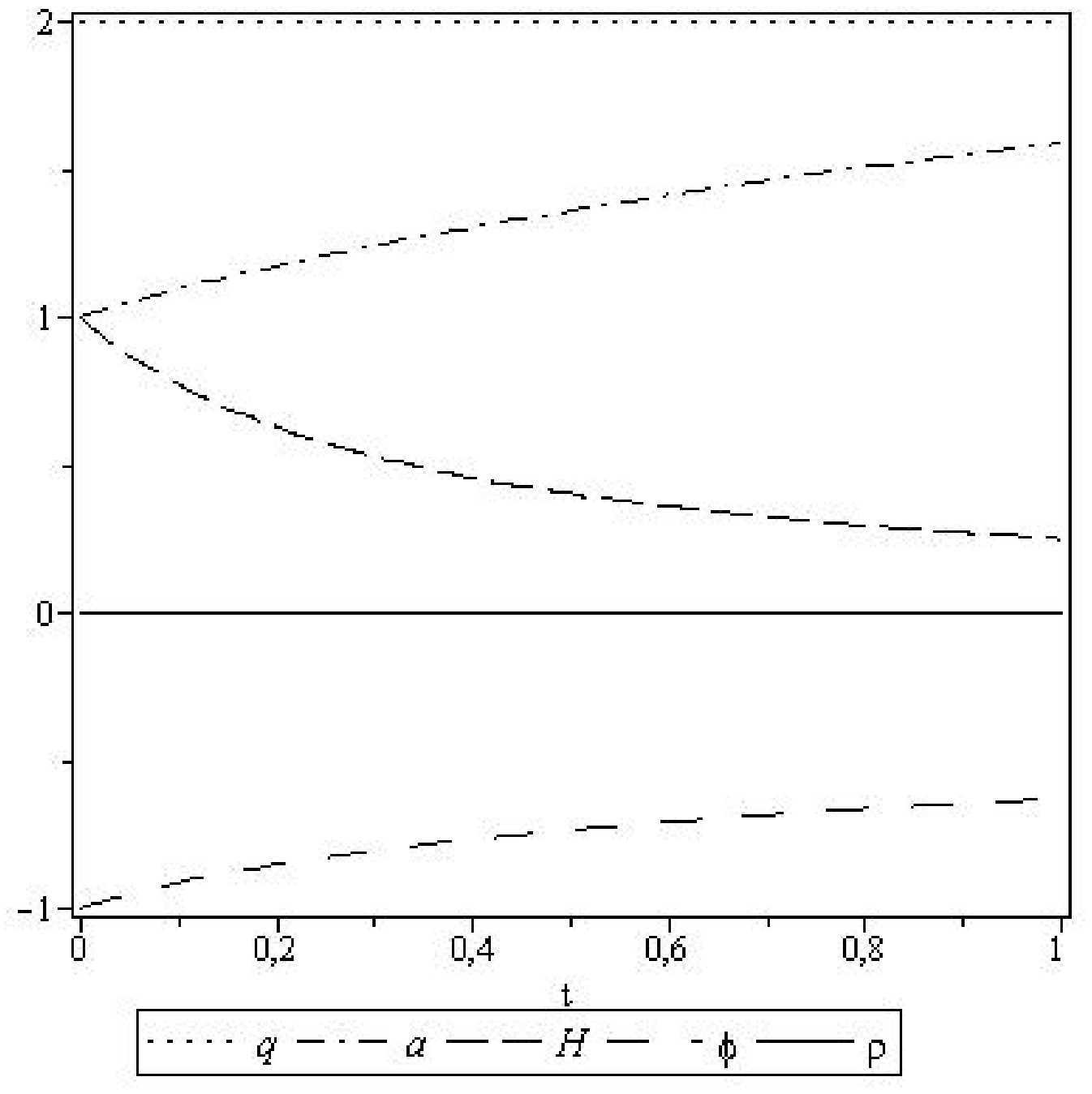}
\includegraphics[scale=0.5]{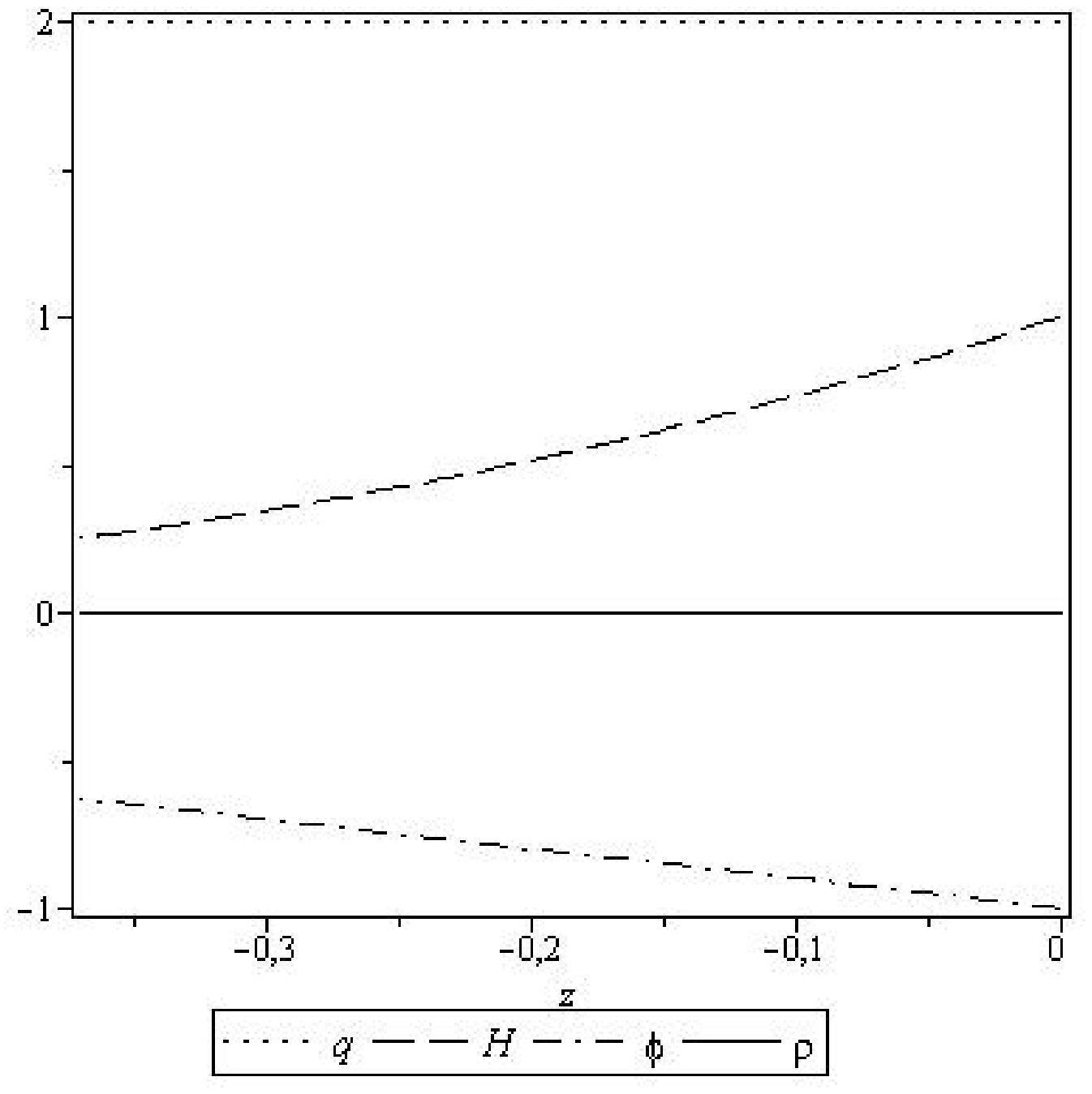}
\caption{$k=0$, $\omega=1$.}
\label{fig1}
\end{figure}

\begin{figure}[!htb]
\includegraphics[scale=0.5]{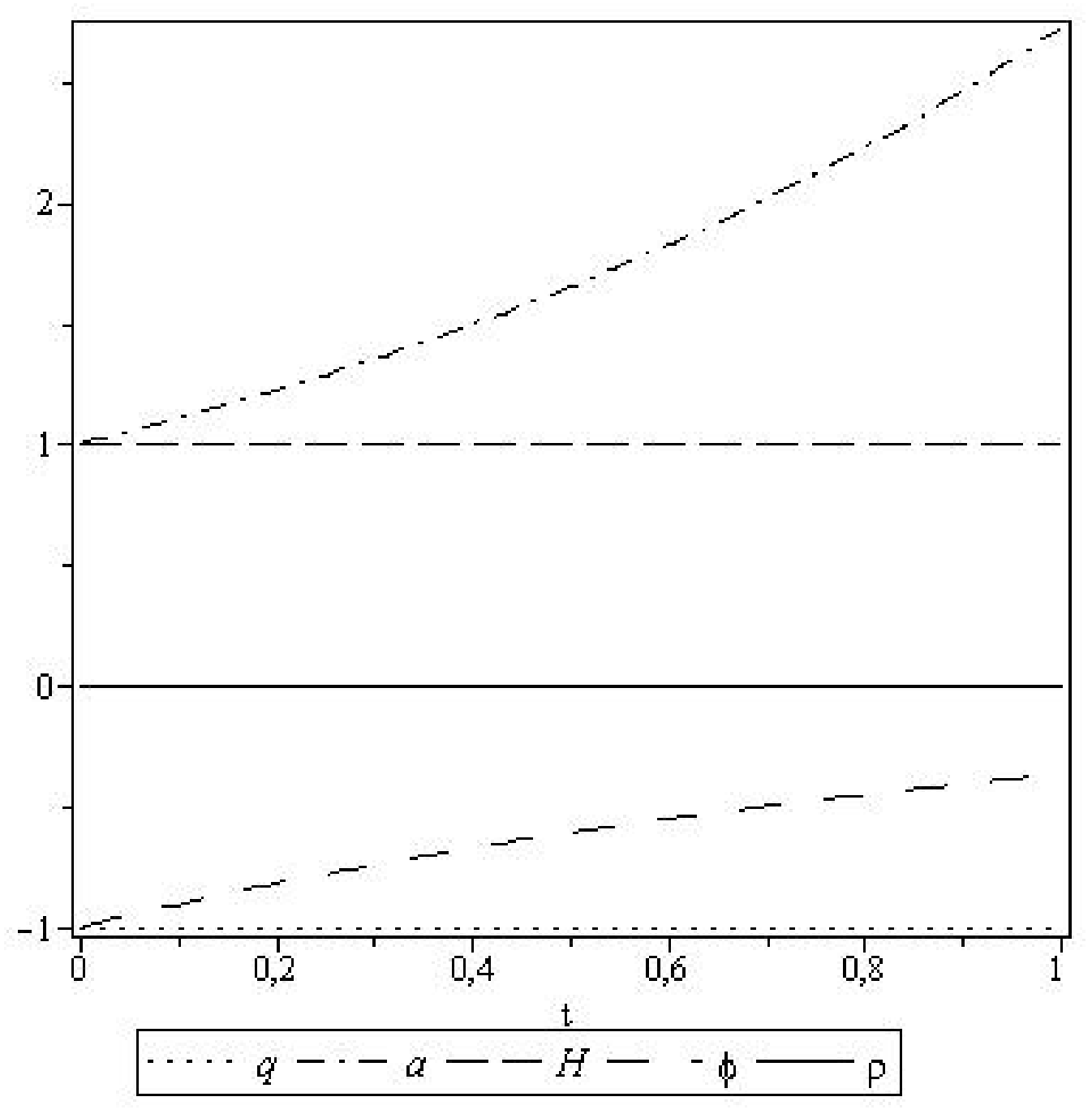}
\includegraphics[scale=0.5]{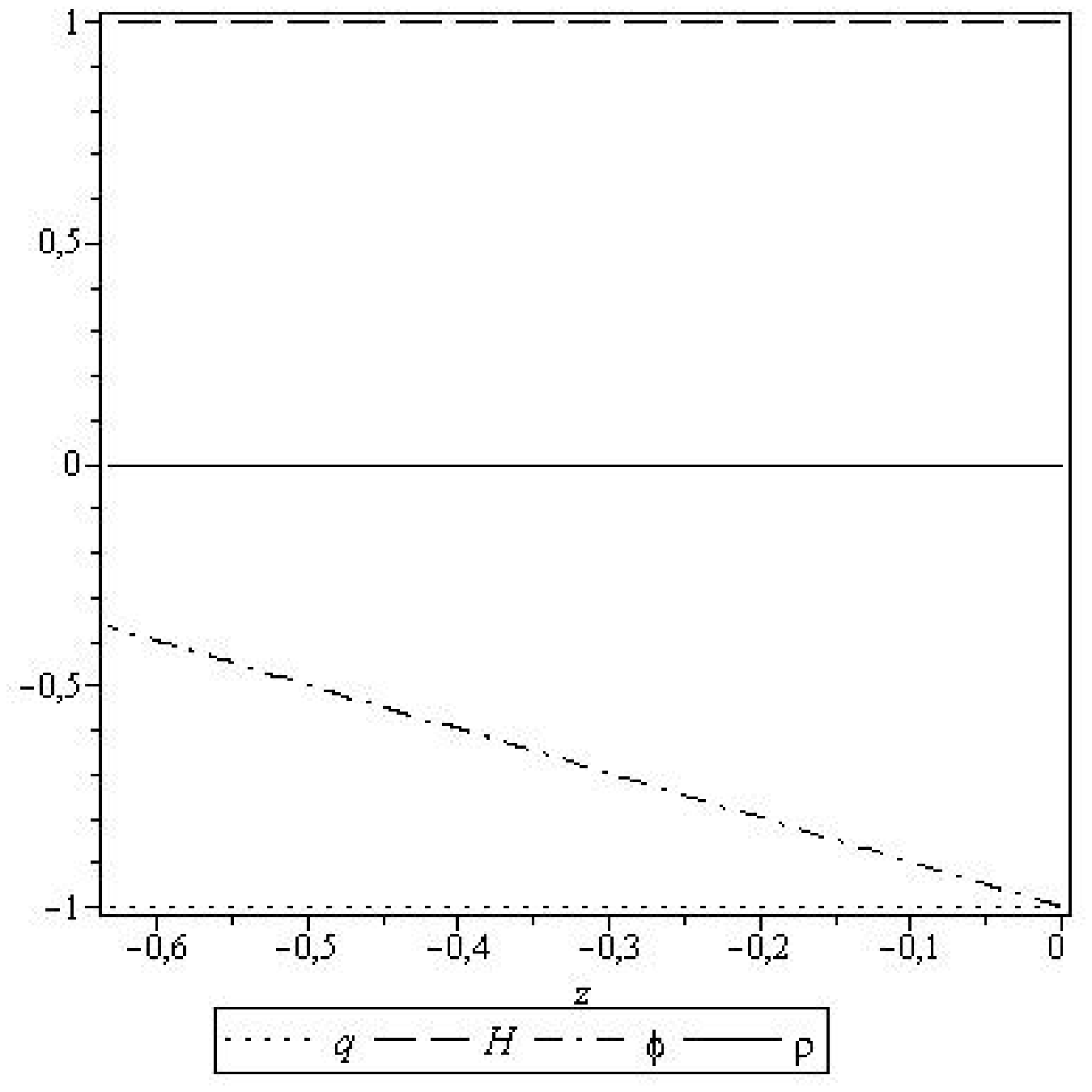}
\caption{$k=0$, $\omega=-1$.}
\label{fig2}
\end{figure}

\begin{figure}[!htb]
\includegraphics[scale=0.5]{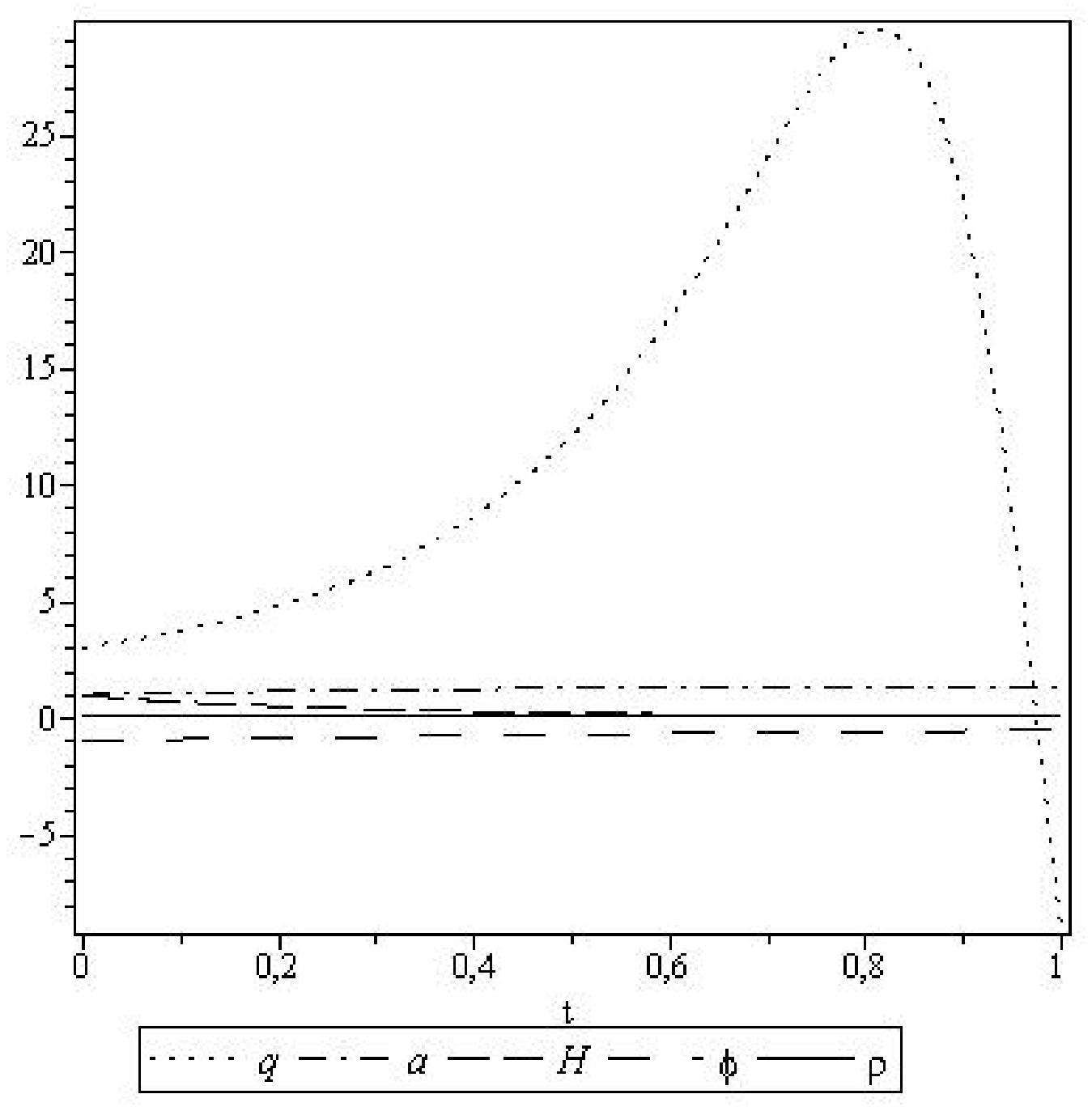}
\includegraphics[scale=0.5]{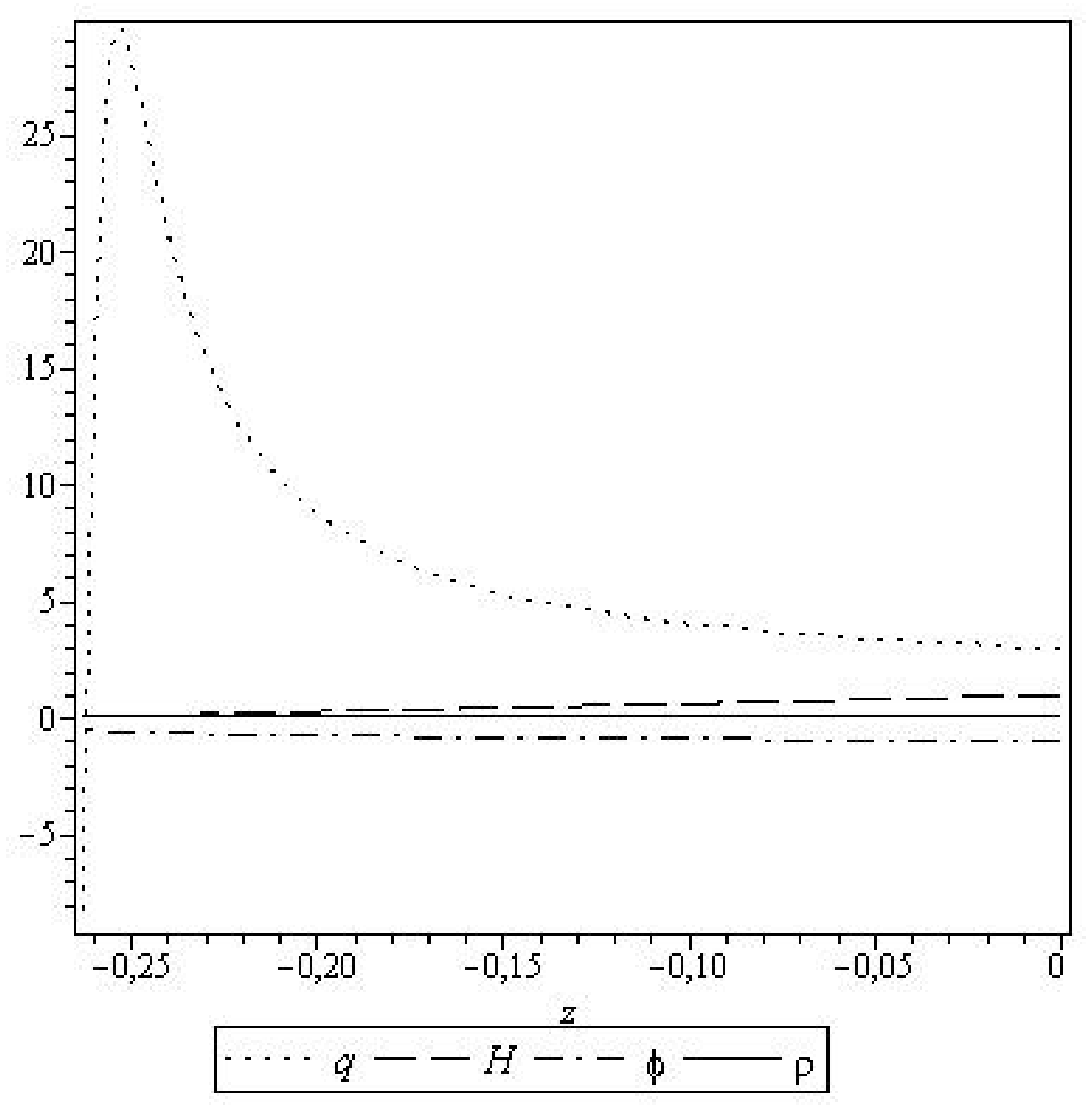}
\caption{$k=1$, $\omega=1$.}
\label{fig3}
\end{figure}

\begin{figure}[!htb]
\includegraphics[scale=0.5]{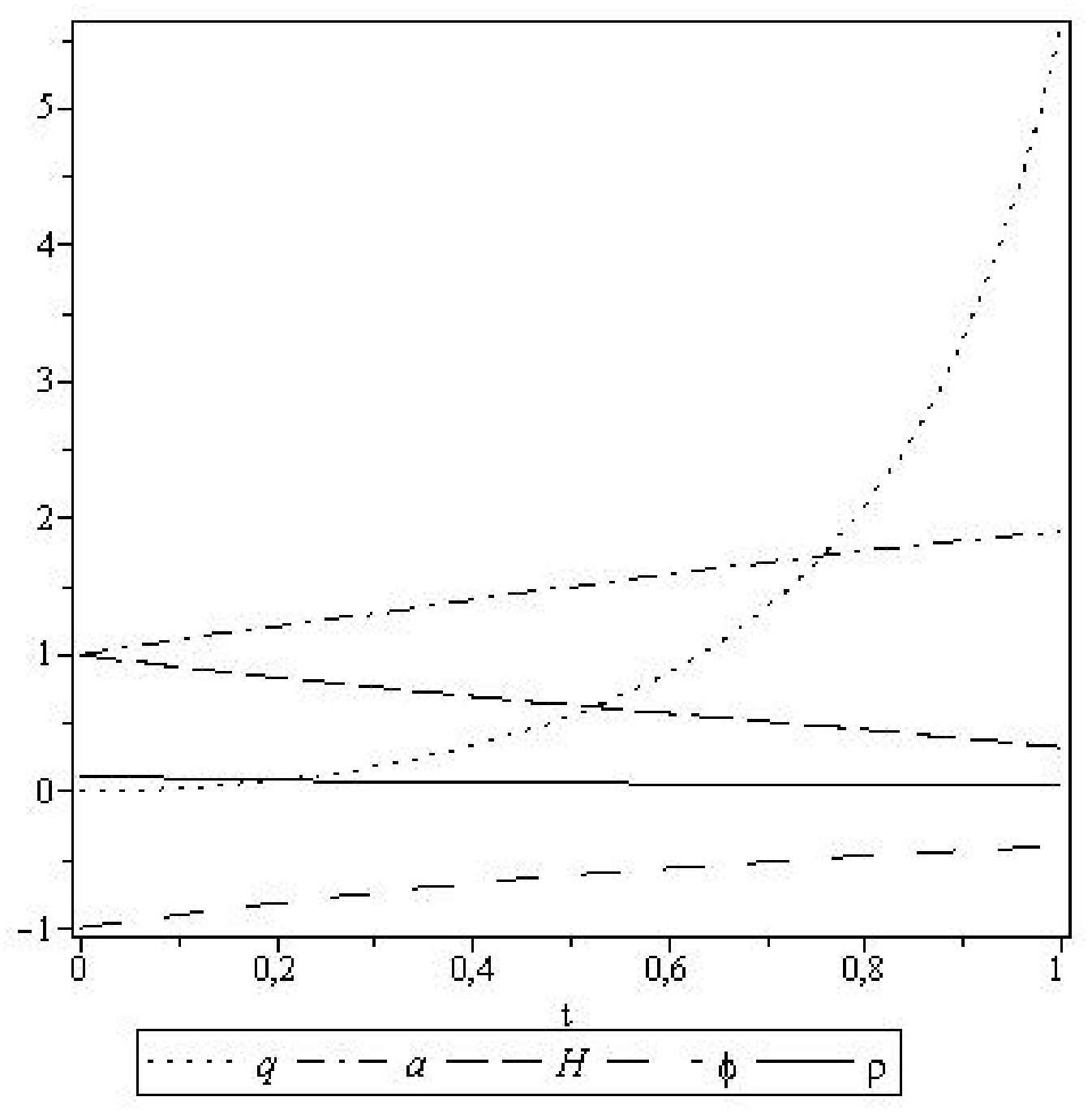}
\includegraphics[scale=0.5]{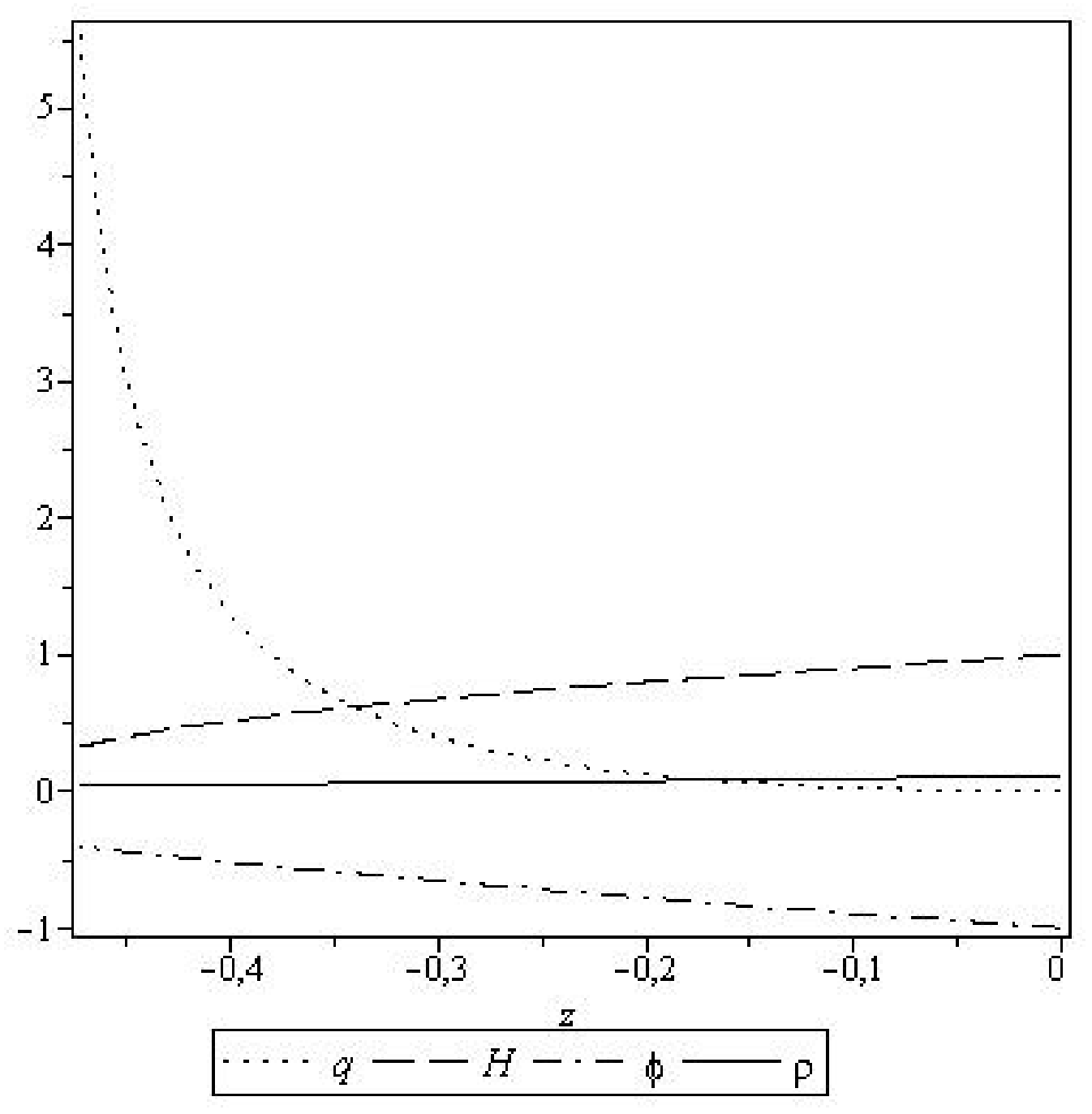}
\caption{$k=1$, $\omega=-1$.}
\label{fig4}
\end{figure}

\begin{figure}[!htb]
\includegraphics[scale=0.5]{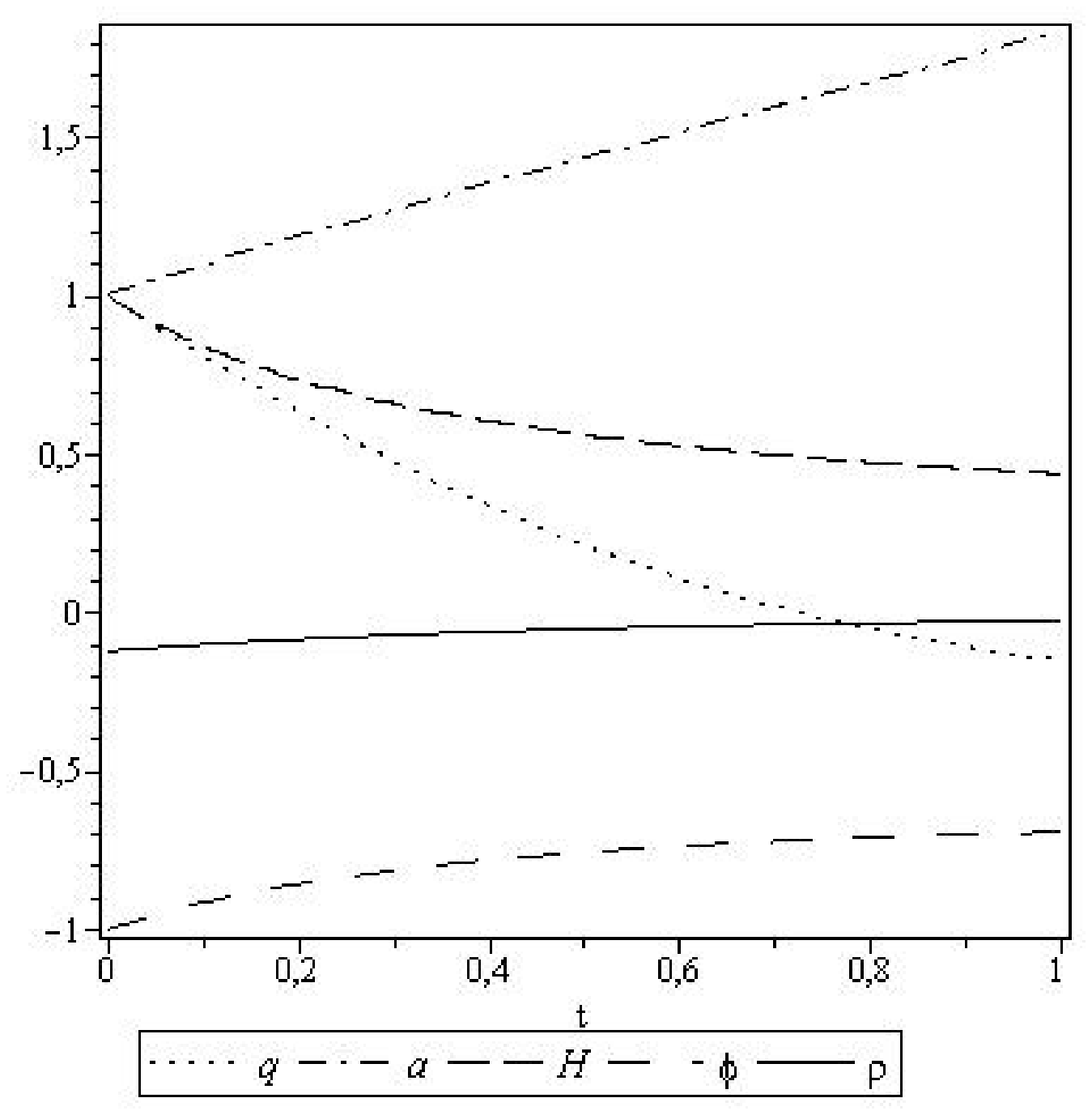}
\includegraphics[scale=0.5]{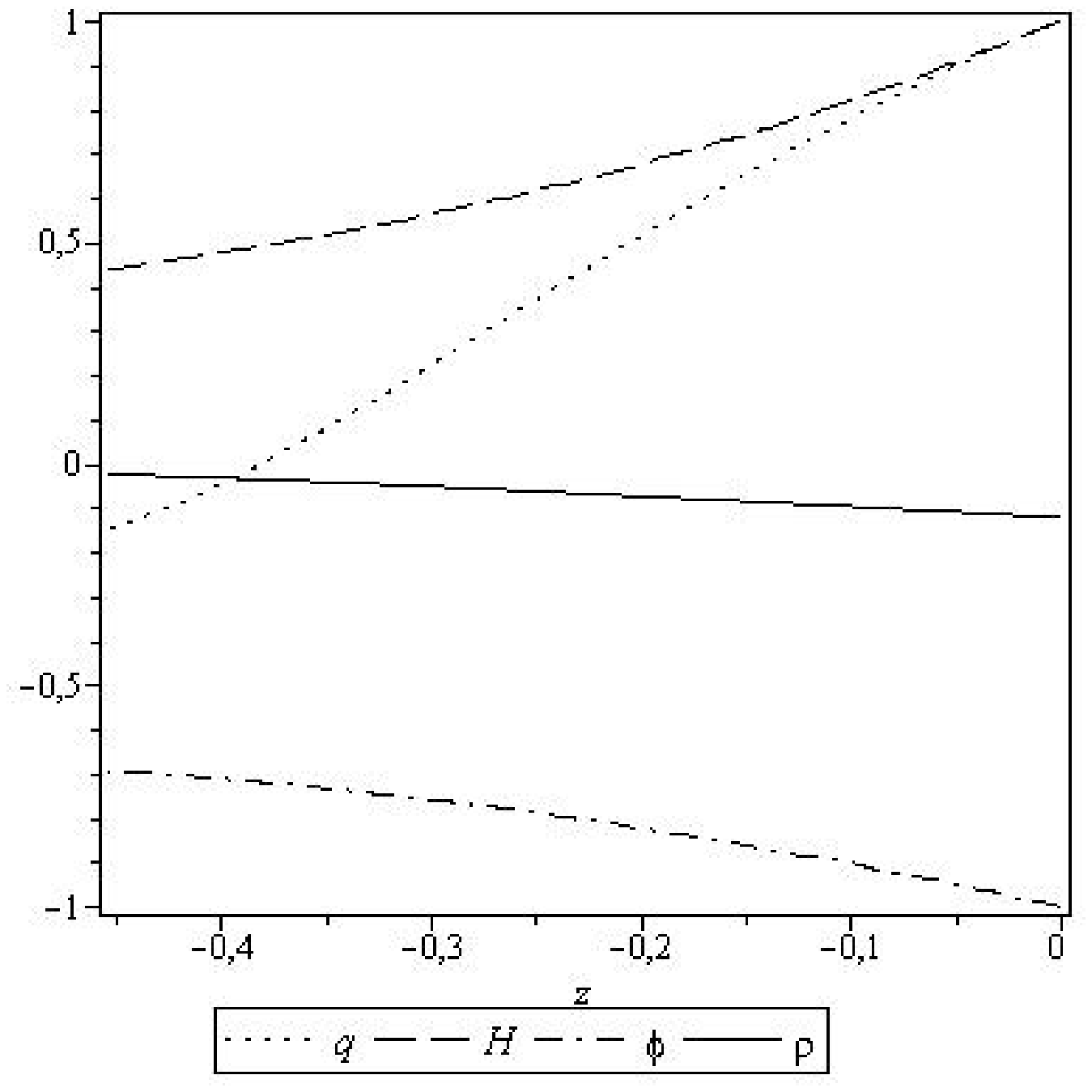}
\caption{$k=-1$, $\omega=1$.}
\label{fig5}
\end{figure}

\begin{figure}[!htb]
\includegraphics[scale=0.5]{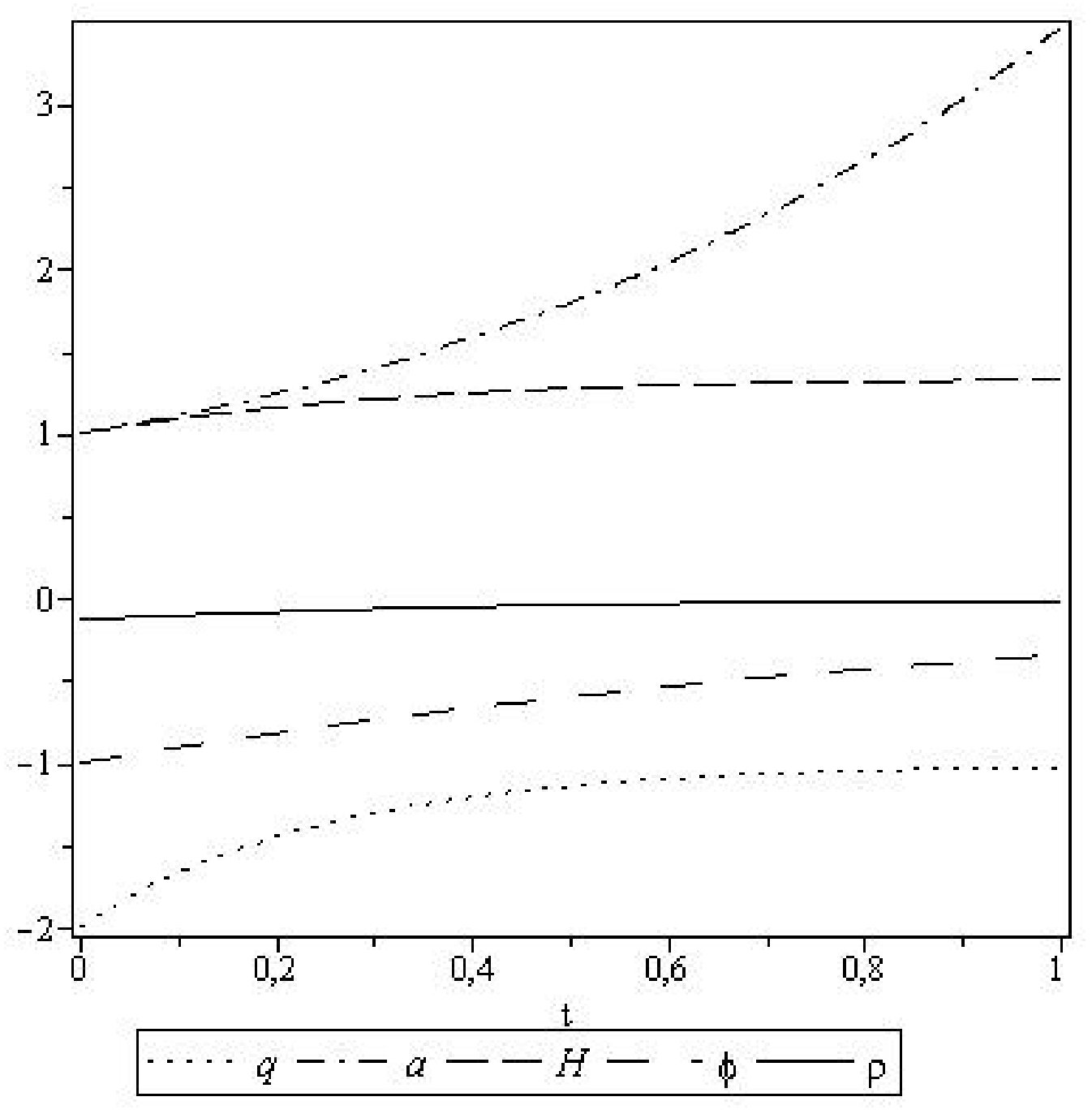}
\includegraphics[scale=0.5]{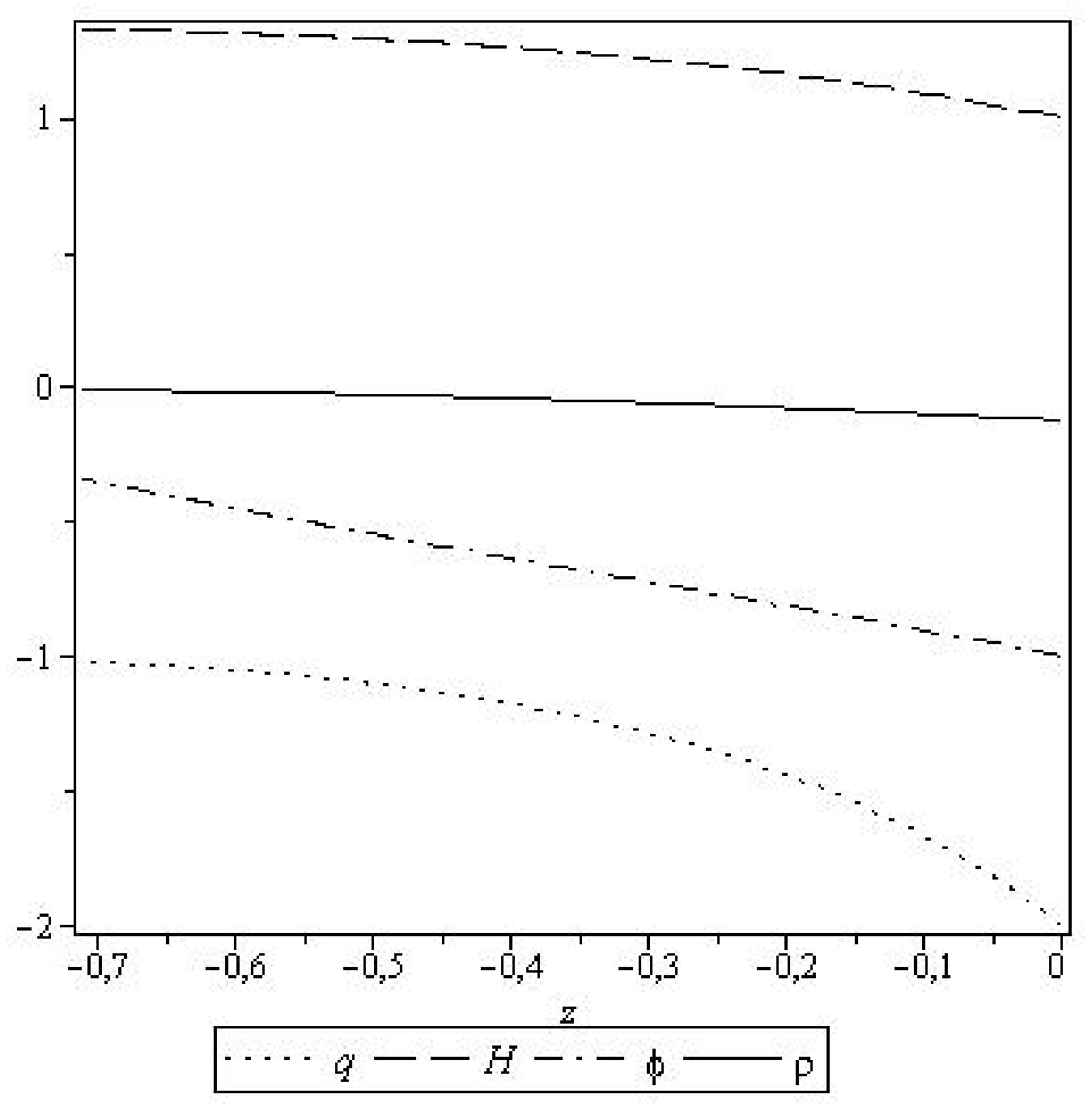}
\caption{$k=-1$, $\omega=-1$.}
\label{fig6}
\end{figure}

Next we solve numerically the field equations. However it should be noted that there are two independent field equations and three unknown variables, since eqs. (\ref{00}) and (\ref{11}) combined yield the trace equation. Thus in order to solve them we have to impose an equation of state for the dark fluid as $p_D=\omega\rho_D$. Such an imposition on the dark fluid is similar of that in general relativity for the perfect fluid. We also use the trace-free equation of state for the perfect fluid of the form $p=\rho/3$. Then
we present our numerical results in figures 1 to 6. Firstly it should be noted that the field equations could be written in terms of H(t), a(t) and $\beta(t)=\frac{\dot{\phi}}{\phi}$, hence it is necessary initial conditions only for those variables. As a natural consequence the model is independent of the choice of $\phi(0)$, however for convenience  we set it as $-1$. We have chosen $H(0)=a(0)=1$ in order to get normalized deviation from those parameters. In this way the redshift defined as $1+z=\frac{a(0)}{a(t)}$ is approximatively given by $z\approx t$ for expansions around $t=0$. We plot the cosmological quantities in terms of the redshift always next to those in terms of time. We excluded from these panels $a(z)$ since it is given by the definition already written. For consistence with field equations $\beta(0)=-1$ for $k=0$, thus we use such a value as a reference for all models. The next interesting feature displayed in the panels that should be noted is the dependence of the cosmological quantities with the model parameter $\omega$. Another feature that should be noted is that the density of the ordinary fluid is slightly different from zero, thus the responsible by any acceleration is the very field $\phi$. From figures \ref{fig1} to \ref{fig2} we see, for $k=0$, two different behaviors. For $\omega=-1$ we note an Universe expanding and accelerating, on the other hand for $\omega=1$ we see a decelerating and expanding Universe. From figures \ref{fig3} to \ref{fig4} we observe the same two behaviors as before for $k=1$. The difference is that for $\omega=1$ there is a point where the Universe experiences an accelerated expansion which does not occur for $\omega=-1$ with just an expansion with deceleration. From figures \ref{fig5} to \ref{fig6} we see for $k=-1$ that there is a change in the deceleration parameter from positive to negative for $\omega=-1$ which indicates a change in the regime of expansion of the Universe. For $\omega=1$ we see only a decelerating and expanding Universe.

\section{Conclusions}\label{con}

In this article we have used the conformal teleparallel gravity to analyze an isotropic and homogeneous Universe. In such a theory the scalar field was introduced in order to have field equations invariant under conformal transformations. Then we have derived an equation for the scalar field in terms of the scalar curvature and the modified Friedmann equations. Such equations have extra terms when compared to the usual ones, thus we associate them to dark energy density and dark pressure, since they drive an acceleration in the Universe. They are written in terms of the scalar field which was introduced in a natural way. We supposed an equation of state for the dark fluid and we solved  numerically the field equations $p=\rho/3$, we also chose $\omega=1$. In addition we worked with the three values of k. We find a deceleration parameter compatible with the observed accelerating Universe at least for a particular choice of the parameters of the theory.

It is important to make clear the difference between our approach and what exist in literature. Thus, as stated before, Conformal Teleparallel gravity has been used to understand cosmological models. In reference \cite{Bamba} the conformal symmetry is introduced by Weyl tensor, we do it by means a scalar field and the modification of the Teleparallel gravity Lagrangian density. Maybe there is a relation between these two approaches in the general case but it is not clear yet. In reference \cite{Momeni} the authors use the same approach as us to deal with conformal symmetry. However the introduction of matter fields is different from our proposal which deals with trace-null energy-momentum tensor. As a consequence we have to assume an equation of state for what we called dark fluid.

It's worth to point out the important role of the field $\phi$ in the accelerated expansion of the Universe. Here the difference between our theory and general relativity is ensured by such field. At the same time $\phi$ is responsible by the conformal symmetry and as an effective density and pressure. This last statement is clear since the density of the fluid, which represents the matter fields, is always very close to zero for any choice of the parameters of our model. Therefore the introduction of conformal symmetry in Teleparallel gravity allows the explanation of an accelerating and expanding Universe.


\section*{Acknowledgements}

This work was partially supported by Conselho Nacional de Desenvolvimento Cient\'ifico e Tecnol\'ogico (CNPq) and Coordena\c{c}\~{a}o de Aperfei\c{c}oamento de Pessoal de N\'{i}vel Superior (CAPES). A.F.S. has been supported by the CNPq project 476166/2013-6 and 201273/2015-2.



\begin{thebibliography}{99}

\makeatletter
\providecommand \@ifxundefined [1]{%
 \@ifx{#1\undefined}
}%
\providecommand \@ifnum [1]{%
 \ifnum #1\expandafter \@firstoftwo
 \else \expandafter \@secondoftwo
 \fi
}%
\providecommand \@ifx [1]{%
 \ifx #1\expandafter \@firstoftwo
 \else \expandafter \@secondoftwo
 \fi
}%
\providecommand \natexlab [1]{#1}%
\providecommand \enquote  [1]{``#1''}%
\providecommand \bibnamefont  [1]{#1}%
\providecommand \bibfnamefont [1]{#1}%
\providecommand \citenamefont [1]{#1}%
\providecommand \href@noop [0]{\@secondoftwo}%
\providecommand \href [0]{\begingroup \@sanitize@url \@href}%
\providecommand \@href[1]{\@@startlink{#1}\@@href}%
\providecommand \@@href[1]{\endgroup#1\@@endlink}%
\providecommand \@sanitize@url [0]{\catcode `\\12\catcode `\$12\catcode
  `\&12\catcode `\#12\catcode `\^12\catcode `\_12\catcode `\%12\relax}%
\providecommand \@@startlink[1]{}%
\providecommand \@@endlink[0]{}%
\providecommand \url  [0]{\begingroup\@sanitize@url \@url }%
\providecommand \@url [1]{\endgroup\@href {#1}{\urlprefix }}%
\providecommand \urlprefix  [0]{URL }%
\providecommand \Eprint [0]{\href }%
\providecommand \doibase [0]{http://dx.doi.org/}%
\providecommand \selectlanguage [0]{\@gobble}%
\providecommand \bibinfo  [0]{\@secondoftwo}%
\providecommand \bibfield  [0]{\@secondoftwo}%
\providecommand \translation [1]{[#1]}%
\providecommand \BibitemOpen [0]{}%
\providecommand \bibitemStop [0]{}%
\providecommand \bibitemNoStop [0]{.\EOS\space}%
\providecommand \EOS [0]{\spacefactor3000\relax}%
\providecommand \BibitemShut  [1]{\csname bibitem#1\endcsname}%
\let\auto@bib@innerbib\@empty

\bibitem{Perlm} S. Perlmutter et al. [SNCP Collaboration], Astrophys. J. {\bf 517}, 565 (1999), arXiv:astro-ph/9812133; A. G. Riess et al. [Supernova Search Team Collaboration], Astron. J. {\bf 116}, 1009 (1998), arXiv:astro-ph/9805201.

\bibitem{Tegmark} M. Tegmark et al. [SDSS Collaboration], Phys. Rev. D {\bf 69}, 103501 (2004), arXiv:astro-ph/0310723; U. Seljak et al. [SDSS Collaboration], ibid. {\bf 71}, 103515 (2005), arXiv:astro-ph/0407372.

\bibitem{Eisenstein} D. J. Eisenstein et al. [SDSS Collaboration], Astrophys. J. {\bf 633}, 560 (2005), arXiv:astro-ph/0501171.

\bibitem{Spergel} D. N. Spergel et al. [WMAP Collaboration], Astrophys. J. Suppl. {\bf 148}, 175 (2003), arXiv:astro-ph/0302209; ibid. {\bf 170}, 377 (2007), arXiv:astro-ph/0603449; E. Komatsu et al. [WMAP Collaboration], ibid. {\bf 180}, 330 (2009), arXiv:0803.0547 [astroph]; ibid. {\bf 192}, 18 (2011), arXiv:1001.4538 [astro-ph.CO]; G. Hinshaw et al., arXiv:1212.5226 [astro-ph.CO]; P. A. R. Ade et al. [Planck Collaboration], arXiv:1303.5076 [astro-ph.CO].

\bibitem{Jain} B. Jain and A. Taylor, Phys. Rev. Lett. {\bf 91}, 141302 (2003), arXiv:astro-ph/0306046.

\bibitem{MG}
 Clifton, Timothy; Ferreira, Pedro G.; Padilla, Antonio; Skordis, Constantinos. Physics Reports, {\bf 513}, 1, p. 1-189, arXiv:1106.2476.

\bibitem{Maluf1} J. W. Maluf and F. F. Faria, Phys. Rev. D {\bf 85}, 027502 (2012), arXiv:1110.3095 [gr-qc].

\bibitem{CEM} C. L\"{u}bbe and J. A. Valiente Kroon, Ann. Phys. {\bf 328}, 1-25 (2013), arXiv:1111.4691v1 [gr-qc].

\bibitem{Maluf2}  J. W. Maluf and F. F. Faria, Annalen Phys. {\bf 524}, 366 (2012), arXiv:1203.0040 [gr-qc].

\bibitem{Ulhoa}
J.~W. Maluf, S.~C. Ulhoa, F.~F. Faria, and J.F.. daRocha Neto.
Class. Quant. Grav. {\bf 23}, 22, 6245-6256 (2006).

\bibitem{maluf:335}
Jose~W. Maluf.
J. Math. Phys. {\bf 35}, 1, 335-343 (1994).

\bibitem{maluf21}
J.~W. Maluf.
Annalen Phys. {\bf 14}, 723-732 (2005).

\bibitem{Matthias} M. R. Gaberdiel, Rept. Prog. Phys. {\bf 63}, 607 (2000), arXiv:hep-th/9910156.

\bibitem{Haw} S.W. Hawking and G.F.R. Ellis, {\it The Large Scale Structure of Space-time} (Cambridge University Press, 1999).

\bibitem{Fujii} Y. Fujii and K.-I. Maeda, {\it The Scalar-Tensor Theory of Gravitation} (Cambridge University Press, 2003).

\bibitem{Carroll} S. M. Carroll, {\it An introduction to general relativity spacetime and geometry} (Addison Wesley, 2004).

\bibitem{Cham1} A. H. Chamseddine and V. Mukhanov, JHEP {\bf 1311}, 135 (2013), arXiv:1308.5410 [astro-ph.CO].

\bibitem{Cham2} A. H. Chamseddine, V. Mukhanov and A. Vikman, arXiv:1403.3961 [astro-ph.CO].

\bibitem{Haw2}  S. W. Hawking, T. Hertog and H. S. Reall, Phys. Rev. D {\bf 63}, 083504 (2001) [hep-th/0010232]. S. Nojiri and S. D. Odintsov, Phys. Lett. B {\bf 484}, 119 (2000) [hep-th/0004097].

\bibitem{Arm} R. Armillis, A. Monin and M. Shaposhnikov, arXiv:1302.5619 [hep-th].

\bibitem{Buch1}  I. L. Buchbinder, S. D. Odintsov and I. L. Shapiro, {\it Effective Action in Quantum Gravity} Bristol, UK: IOP (1992) 413 p.

\bibitem{Buch2} I. L. Buchbinder and S. D. Odintsov, Sov. J. Nucl. Phys. {\bf 40}, 848 (1984).

\bibitem{Man1}  P. D. Mannheim, Prog. Part. Nucl. Phys. {\bf 56}, 340 (2006).

\bibitem{Man2} P. D. Mannheim, arXiv:1101.2186.

\bibitem{Moon} T. Moon, P. Oh and J. Sohn, JCAP {\bf 1011}, 005 (2010) [arXiv:1002.2549].

\bibitem{Hooft1} G. ât Hooft, arXiv:1009.0669.

\bibitem{Hooft2} G. ât Hooft, arXiv:1011.0061; Found. Phys. {\bf 41}, 1829 (2011), arXiv:1104.4543.

\bibitem{Formiga} J.B. Formiga, J.B. Fonseca-Neto and C. Romero, Phys. Rev. D {\bf 87} 6, 067702 (2013), arXiv:1302.0900 [gr-qc].

\bibitem{Bamba} K. Bamba, S. D. Odintsov, D. S\'aez-G\'omez, Phys. Rev. D {\bf 88}, 084042 (2013), arXiv:1308.5789 [gr-qc].

\bibitem{Myr} D. Momeni, R. Myrzakulov, Eur. Phys. J. Plus {\bf 129}, 137 (2014), arXiv:1404.0778 [gr-qc].

\bibitem [{\citenamefont {Einstein}(1930)}]{einstein}%
  \BibitemOpen
  \bibfield  {author} {\bibinfo {author} {\bibfnamefont {A.}~\bibnamefont
  {Einstein}},\ }\href@noop {} {\bibfield  {journal} {\bibinfo  {journal}
  {Math. Annal.}\ }\textbf {\bibinfo {volume} {102}},\ \bibinfo {pages} {685}
  (\bibinfo {year} {1930})}\BibitemShut {NoStop}%
\bibitem [{\citenamefont {{Cartan}}(1980)}]{Cartan}%
  \BibitemOpen
  \bibfield  {author} {\bibinfo {author} {\bibfnamefont {E.}~\bibnamefont
  {{Cartan}}},\ }in\ \href@noop {} {\emph {\bibinfo {booktitle} {NATO ASIB
  Proc. 58: Cosmology and Gravitation: Spin, Torsion, Rotation, and
  Supergravity}}},\ \bibinfo {editor} {edited by\ \bibinfo {editor}
  {\bibfnamefont {P.~G.}\ \bibnamefont {{Bergmann}}}\ and\ \bibinfo {editor}
  {\bibfnamefont {V.}~\bibnamefont {{de Sabbata}}}}\ (\bibinfo {year} {1980})\
  pp.\ \bibinfo {pages} {489--491}\BibitemShut {NoStop}%
\bibitem [{\citenamefont {de~Andrade}\ \emph {et~al.}(2000)\citenamefont
  {de~Andrade}, \citenamefont {Guillen},\ and\ \citenamefont
  {Pereira}}]{PhysRevLett.84.4533}%
  \BibitemOpen
  \bibfield  {author} {\bibinfo {author} {\bibfnamefont {V.~C.}\ \bibnamefont
  {de~Andrade}}, \bibinfo {author} {\bibfnamefont {L.~C.~T.}\ \bibnamefont
  {Guillen}}, \ and\ \bibinfo {author} {\bibfnamefont {J.~G.}\ \bibnamefont
  {Pereira}},\ }\href {\doibase 10.1103/PhysRevLett.84.4533} {\bibfield
  {journal} {\bibinfo  {journal} {Phys. Rev. Lett.}\ }\textbf {\bibinfo
  {volume} {84}},\ \bibinfo {pages} {4533} (\bibinfo {year}
  {2000})}\BibitemShut {NoStop}%
\bibitem{Landau}
 \textsc{L.\,D. Landau} and  \textsc{E.\,M. Lifshiz},
The Classical Theory of Fields, 4th revised english edition, Course of
  Theoretical Physics,  Vol.\,2 (Elsevier Butterworth-Heinemann, 2004).
 \bibitem{Maluf:2007qq}
 J.\,W. Maluf, F.\,F. Faria and S.\,C. Ulhoa. Class. Quant. Grav. \textbf{24}, 2743-2754 (2007).
\bibitem{Momeni}
Davood Momeni and Ratbay Myrzakulov. Eur. Phys. J. Plus {\bf 129}, 6, 137 (2014).
\end{thebibliography}
\end{document}